\documentclass[bib]{ba}

\pdfoutput=1

\usepackage{bayes}

\usepackage{amsmath,amsfonts,amsthm,amssymb,latexsym}
\usepackage{graphicx}
\usepackage{natbib}

\newtheorem{example}{Example}
\newcounter{exA}

\newcommand\labelresdataset\relax

\newcommand{\bx}{\mathbf{x}}
\newcommand{\btheta}{\boldsymbol{\theta}}
\newcommand\figsigmc\relax
\newcommand\figtoy\relax
\newcommand\figsismix\relax
\newcommand\figbetabinpmc\relax
\newcommand\figsispmco\relax
\newcommand\figsispmct\relax
\newcommand\dropcap\noindent

\begin{document}

\inserttype[ba0001]{article}
\author{Grelaud, Marin, Robert, Rodolphe and Taly}{
Aude Grelaud\footnote{aude.grelaud@jouy.inra.fr}\\
INRA Jouy-en-Josas, MIG, CEREMADE, Universit\'e Paris Dauphine\\ \& CREST, INSEE, France
\and
 Christian P. Robert\footnote{xian@ceremade.dauphine.fr}\\
CEREMADE, Universit\'e Paris Dauphine \& CREST, INSEE, France
\and
Jean-Michel Marin\footnote{jean-michel.marin@univ-montp2.fr}\\
Institut de Math\'ematiques et Mod\'elisation de Montpellier, Universit\'e Montpellier 2\\
\& CREST, INSEE, France 
\and
Fran\c cois Rodolphe\footnote{francois.rodolphe@jouy.inra.fr}$\quad$
and$\quad$ Jean-Fran\c cois Taly\footnote{jean-francois.taly@jouy.inra.fr}\\
INRA Jouy-en-Josas, MIG, France
}
\title[ABC model choice]{ABC likelihood-free methods for model choice in Gibbs random fields}

\maketitle

\begin{abstract}
Gibbs random fields (GRF) are polymorphous statistical models that can be used to analyse different types of dependence,
in particular for spatially correlated data. However, when those models are faced with the challenge
of selecting a dependence structure 
from many, the use of standard model choice methods is hampered by the unavailability of the normalising constant in
the Gibbs likelihood. In particular, from a Bayesian perspective, the computation of the posterior probabilities 
of the models under competition requires special likelihood-free simulation techniques like the 
Approximate Bayesian Computation (ABC) algorithm that is intensively used in population genetics.
We show in this paper how to implement an ABC algorithm geared towards model choice in the general setting of 
Gibbs random fields, demonstrating in particular that there exists a sufficient statistic across models. The accuracy
of the approximation to the posterior probabilities can be further improved by importance sampling on the distribution
of the models. The practical aspects of the method are detailed through two applications, the test of an iid Bernoulli
model versus a first-order Markov chain, and the choice of a folding structure for two proteins.
\end{abstract}


\section{Introduction}

\subsection{Gibbs random fields}

We consider a finite set of sites $\mathcal{S}=\{1,\cdots,n\}$. At each site $i \in \mathcal{S}$, we observe $x_i \in \mathcal{X}_i$ where $\mathcal{X}_i$
is a finite set of states. $\mathcal{X}=\prod_{i=1}^n\mathcal{X}_i$ is the set of the configurations, $\bx=(x_1,\cdots,x_n)$ corresponding to one configuration.
We also consider an undirected graph $\mathcal{G}=(E(\mathcal{G}),V(\mathcal{G}))$ on $\mathcal{S}$, $V(\mathcal{G})$ being a vertex set and $E(\mathcal{G})$ an edge set. 
The sites $i$ and $i^{'}$ are said neighbours (denoted $i\sim i^{'}$) if $(i,i^{'}) \in E(\mathcal{G})$, in other words, if there is a vertex between $i$ and $i^{'}$.
A clique $c$ is a subset of $\mathcal{S}$ where all elements are mutual neighbours \citep{Da80}. We denote by $\mathcal{C}$ the set of all cliques of the undirected
graph $\mathcal{G}$.

In the finite framework previously adopted, Gibbs Random Fields (GRFs) are probabilistic models associated with densities
(with respect to the counting measure) 
\begin{equation}\label{Gibbs}
f(\bx)=\dfrac{1}{Z}\exp\{-U(\bx)\}=\dfrac{1}{Z}\exp\left\{-\sum_{c\in \mathcal{C}}U_c(\bx)\right\}\,,
\end{equation}
where $U(\bx)=\sum_{c\in \mathcal{C}}U_c(\bx)$ is the potential and $Z$ is the corresponding normalising constant 
$$
Z=\sum_{\bx\in\mathcal{X}}\exp\left\{-\sum_{c\in \mathcal{C}}U_c(\bx)\right\}\,.
$$
If the density $f$ of a Markov Random Field (MRF) is everywhere positive, then the Hammersley-Clifford theorem establishes
that there exists a GRF representation of this MRF \citep{Be74}.

We consider here GRF with potential $U(\bx)=-\btheta^\text{T} S(\bx)$
where $\btheta\in\mathbb{R}^p$ is a scale parameter, $S(\cdot)$ is a function taking values in $\mathbb{R}^p$.
$S(\bx)$ is defined on the cliques of the neighbourhood system in that $S(\bx)=\sum_{c\in \mathcal{C}}S_c(\bx)$. 
In that case, we have
\begin{equation}\label{Gee}
f(\bx|\btheta)=\dfrac{1}{Z_{\btheta}}\exp\{\btheta^\text{T}S(\bx)\}\,,
\end{equation}
the normalising constant $Z_{\btheta}$ now depends on the scale parameter $\btheta$.

\subsection{Bayesian model choice}

When considering model selection within this class of Gibbs models, the primary difficulty to address is the unavailability of
the normalising constant $Z_{\btheta}$.  In most realistic settings, the summation
$$
Z_{\btheta} = \sum_{\bx\in\mathcal{X}} \exp\{\btheta^\text{T}S(\bx)\}
$$
involves too many terms to be manageable. Numerical approximations bypassing this constant like path sampling \citep{Ge98}, 
pseudo likelihood \citep{Be75} or those based on an auxiliary variable \citep{Mo06} are not always available either because they
require heavy computations or because they are not accurate enough in the case of the pseudo-likelihood. In particular, selecting
a model with sufficient statistic $S_0$ taking values in $\mathbb{R}^{p_0}$ versus a model with sufficient statistics $S_1$
taking values in $\mathbb{R}^{p_1}$ relies on the Bayes factor corresponding to
the priors $\pi_0$ and $\pi_1$ on the respective parameter spaces
$$
BF_{m_0/m_1}(\bx) = \int \exp\{\btheta_0^\text{T} S_0(\bx)\} / Z_{\btheta_0,0} \pi_0(\text{d}\btheta_0) \bigg/
$$
$$
         \int \exp\{\btheta_1^\text{T}S_1(\bx)\} / Z_{\btheta_1,1} \pi_1(\text{d}\btheta_1)
$$
but this quantity is not easily computable. One faces the same computational difficulties with the posterior probabilities of the models since they also
depend on those unknown constants. To properly approximate those posterior quantities, we thus propose an alternative resolution based on  likelihood-free 
techniques such as Approximate Bayesian Computation (ABC) \citep{Pr99} and we show how ABC is naturally tuned for this purpose by providing a direct
estimator of the Bayes factor.


>From a modelling perspective, GRF are used to model the dependency 
within spatially correlated data, with applications in epidemiology 
\citep{Ge02} and  image analysis \citep{Ib03}, among
others \citep{Ru05}. 
They often use a Potts model defined by a sufficient statistic $S$ taking values in $\mathbb{R}$ in that
$$
S(\bx)=\sum_{i' \sim i} \mathbb{I}_{\{x_i=x_{i'}\}}\,,
$$
where  $\sum_{i' \sim i}$ indicates that the summation is taken over all the neighbour pairs. In that case,  $\mathcal{X}=\{1,\cdots ,K \}^{n}$, $K=2$ corresponding to the Ising model, and $\theta$ is a scalar. 
$S(\cdot)$ therefore monitors the number of identical neighbours over $\mathcal{X}$. 

\subsection{Plan}

For a fixed neighbourhood or model, the unavailability of $Z_{\theta}$ complicates inference
on the scale parameter $\theta$, but the difficulty is increased manifold when several neighbourhood structures
are under comparison. In section \ref{sec:methods}, we describe the main likelihood-free algorithms before proposing a procedure based on an ABC algorithm aimed at selecting a model. Then, we show how to improve the accuracy of this approximation using an importance sampling procedure. In section \ref{sec:res}, we consider the toy example
of an iid sequence [with trivial neighbourhood structure] tested against a Markov chain model [with nearest neighbour structure]
as well as a biophysical example aimed at selecting a protein 3D structure.

\section{Methods}
\label{sec:methods}
\subsection{Approximate Bayesian Computation}
When the likelihood is not available in closed form, there exist likelihood-free methods that overcome the
difficulty faced by standard simulation techniques via a basic acceptance-rejection algorithm. The algorithm on
which the ABC method [introduced by \cite{Pr99} and expanded in \cite{Be02} and \cite{Mar03}] is based can be briefly 
described as follows: given a dataset $\bx^0=(x_1,\cdots,x_n)$ associated with the sampling distribution $f(\cdot|\theta)$,
and under a prior distribution $\pi(\theta)$ on the parameter $\theta$, this method generates a parameter value from
the posterior distribution $\pi(\theta|\bx^0)\propto\pi(\theta) f(\bx^0|\theta)$ by simulating jointly a value $\theta^{*}$
from the prior, $\theta^{*}\sim\pi(\cdot)$, and a value $\bx^*$ from the sampling distribution $\bx^*\sim f(\cdot|\theta^*)$ until
$\bx^*$ is equal to the observed dataset $\bx^0$. The rejection algorithm thus reads as\\

\hrule
\smallskip
\noindent{\em Exact rejection algorithm: }
\begin{sffamily}
\begin{enumerate}
 \item Generate $\theta^{*}$ from the prior $\pi$.
 \item Generate $\bx^{*}$ from the model $f(\cdot|\theta^{*})$. 
 \item Accept $\theta^{*}$ if $\bx^{*}=\bx^0$, otherwise, start again in 1.
\end{enumerate}
\end{sffamily}
\hrule
\medskip
This solution is not approximative in that the output is truly simulated from the posterior distribution $\pi(\theta|\bx^0)$
$\propto f(\bx^0|\theta)\pi(\theta)$ since $(\theta^{*}, \bx^{*})\sim \pi(\theta^{*}) \mathbb{I}_{\{\bx^*=\bx^0\}}f(\bx^*|\theta)$.
In many settings, including those with continuous observations $\bx^0$, it is however impractical or
impossible  to wait for $\bx^{*}=\bx^0$ to occur and the approximative solution
is to introduce a tolerance in the test, namely to accept $\theta^*$ if simulated data and observed data are close enough, in the sense of
a distance $\rho$, given a fixed tolerance level $\epsilon$,  $\rho(\bx^{*},\bx^0)<\epsilon$. The distance $\rho$ is open to choice but is usually
an Euclidean distance $\rho(\bx^{*},\bx^0)=\sum_{i=1}^n (x_i^*-x_i^0)^2$ (see \cite{Be02} or \cite{Bl08}). The corresponding $\epsilon$-tolerance rejection algorithm is then\\
\hrule
\smallskip
\noindent{\em $\epsilon$-tolerance rejection algorithm: }
\begin{sffamily}
\begin{enumerate}
 \item Generate $\theta^{*}$ from the prior $\pi$.
 \item Generate $\bx^{*}$ from the model $f(\cdot|\theta^{*})$.
 \item Accept $\theta^{*}$ if $\rho(\bx^{*},\bx^0)<\epsilon$, otherwise, start again in 1.
\end{enumerate}
\end{sffamily}
\hrule
\medskip
This approach is obviously approximative when $\epsilon\ne 0$.
The output from the $\epsilon$-tolerance rejection algorithm is thus associated 
with the distribution 
$$
\pi(\theta|\rho(\bx^{*},\bx^0)<\epsilon)\propto \pi(\theta)\mathbb{P}_{\theta}(\rho(\bx^{*},\bx^0)<\epsilon)
$$
with $\mathbb{P}_{\theta}(\rho(\bx^{*},\bx^0)<\epsilon)=\int \mathbb{I}_{\left\lbrace \rho(\bx^{*},\bx^0)<\epsilon\right\rbrace }f(\bx^*|\theta^{*})\text{d}\bx^*$.
The choice of $\epsilon$ is therefore paramount for good performances of the method. 
If $\epsilon$ is too large, the approximation is poor; when $\epsilon\rightarrow\infty$, it amounts to simulating
from the prior since all simulations are accepted (as $\mathbb{P}_{\theta}(\rho(\bx^{*},\bx^0)<\epsilon)\rightarrow 1 \mbox{ when }\epsilon \rightarrow \infty$).
If $\epsilon$ is sufficiently small, $\pi(\theta|\rho(\bx^{*},\bx^0)<\epsilon)$ is a good approximation of $\pi(\theta|\bx^0)$.
There is no approximation when $\epsilon= 0$, since the $\epsilon$-tolerance rejection algorithm corresponds to the exact rejection algorithm, 
but the acceptance probability may be too low to be practical. Selecting the ``right'' $\epsilon$ is thus crucial. It is customary to pick $\epsilon$ as 
an empirical quantile of $\rho(\bx^{*},\bx^0)$ when $\bx^*$ is simulated from the marginal distribution
$\bx^*\propto \int \pi(\theta)\mathbb{P}_{\theta}(\rho(\bx^{*},\bx^0)<\epsilon)\text{d}\theta$, and the choice is often the corresponding $1\%$ quantile
(see, for instance \cite{Be02} or \cite{Bl08}). \cite{Wi08} propose to replace the approximation by an exact simulation based on a convolution with an arbitrary kernel. 

The data $\bx^0$ usually being of a large dimension, another level of approximation is enforced within the true ABC algorithm, by replacing the
distance $\rho(\bx^{*},\bx^0)$ with a corresponding distance between summary statistics $\rho(S(\bx^{*}),S(\bx^0))$ \citep{Be02}.
When $S$ is a sufficient statistic, this step has no impact on the approximation since $\pi(\theta|\rho(S(\bx^{*}),S(\bx^0))=\pi(\theta|\rho(\bx^{*},\bx^0)$. In practice, it is rarely the case that a sufficient
statistic of low dimension is available when implementing ABC (see \cite{Be02} or \cite{Bl08}). As it occurs, the setting of model choice among Gibbs random fields is an exception in that it allows
for such a beneficial structure, as will be shown below. In the general case, the output of the ABC algorithm is therefore a simulation from
the distribution $\pi(\theta|\rho(S(\bx^{*}),S(\bx^0))<\epsilon)$. The algorithm reads as follows:\\
\hrule
\smallskip
\noindent{\em ABC algorithm: }
\begin{sffamily}
\begin{enumerate}
 \item Generate $\theta^{*}$ from the prior $\pi$.
 \item Generate $\bx^{*}$ from the model $f(\cdot|\theta^{*})$.
 \item Compute the distance $\rho(S(\bx^0),S(\bx^{*}))$. 
 \item Accept $\theta^{*}$ if $\rho(S(\bx^0),S(\bx^{*}))<\epsilon$, otherwise, start again in 1.
\end{enumerate}
\end{sffamily}
\hrule
\smallskip

\subsection{Model choice via  ABC}\label{sec:model choice}
In a model choice perspective, we face $M$ Gibbs random fields in competition, 
each model $m$ being associated with sufficient statistic $S_m$ $(0\le m\le M-1)$, i.e.~with corresponding likelihood
$$
f_m(\bx|\theta_m)=\exp \left\{ \theta_m^\text{T} S_m(\bx) \right\} \big/ Z_{\theta_m,m}\,,
$$
where $\theta_m\in\Theta_m$ and $Z_{\theta_m,m}$ is the unknown normalising constant.
Typically, the choice is between $M$ neighbourhood relations $i\overset{m}{\sim} i'$ $(0\le m\le M-1)$ with
$S_m(\bx)=\sum_{i\overset{m}{\sim} i'} \mathbb{I}_{\{x_i=x_{i'}\}}$.
>From a Bayesian perspective, the choice between those models is driven by the posterior probabilities of
the models. Namely, if we consider an extended parameter space
$\Theta=\cup_{m=0}^{M-1}\{m\}\times\Theta_{m}$ that includes the model index $\mathcal{M}$,
we can define a prior distribution on the model index $\pi(\mathcal{M}=m)$ as well as a prior 
distribution on the parameter conditional on the value $m$ of the model index, $\pi_m(\theta_m)$, defined on the parameter 
space $\Theta_m$. The computational target is thus the model posterior probability 
$$
\mathbb{P}(\mathcal{M}=m|\bx)\propto\int_{\Theta_m} f_m(\bx|\theta_m) \pi_m(\theta_m) \,\text{d}\theta_m\,\pi(\mathcal{M}=m)\,,
$$
i.e.~the marginal of the posterior distribution on $(\mathcal{M},\theta_0,\ldots,\theta_{M-1})$ given $\bx$.
Therefore, if $S(\bx)$ is a sufficient statistic for the joint parameters $(\mathcal{M},\theta_0,\ldots,\theta_{M-1})$,
$$
\mathbb{P}(\mathcal{M}=m|\bx)=\mathbb{P}(\mathcal{M}=m|S(\bx))\,.
$$

Each model has its own sufficient statistic $S_m(\cdot)$. Then, for each model, the 
vector of statistics $S(\cdot)=(S_0(\cdot),\ldots,S_{M-1}(\cdot))$ is obviously sufficient
(since it includes the sufficient statistic of each model). Moreover, the structure of the Gibbs random
field allows for a specific factorisation of the distribution $f_m(\bx|\theta_m)$. Indeed,
the distribution of $\bx$ in model $m$ factorises as
\begin{eqnarray*}
f_m(\bx|\theta_m) & = & h_m(\bx|S(\bx))g_m(S(\bx)|\theta_m) \\
                  & = & \dfrac{1}{n(S(\bx))}g_m(S(\bx)| \theta_m)
\end{eqnarray*}
where $g_m(S(\bx)|\theta_m)$ is the distribution of $S(\bx)$ within model $m$
[not to be confused with the distribution of $S_m(\bx)$] and where
$$
n(S(\bx))=\sharp\left\{\tilde\bx\in\mathcal{X}:S(\tilde\bx)=S(\bx) \right\}
$$ 
is the cardinality of the set of elements of $\mathcal{X}$ with the same sufficient statistic, which
does not depend on $m$ (the support of $f_m$ is constant with $m$). The statistic $S(\bx)$ is 
therefore also sufficient for the joint parameters $(\mathcal{M},\theta_0,\ldots,\theta_{M-1})$.
That the concatenation of the sufficient statistics of each model is also
a sufficient statistic for the joint parameters $(\mathcal{M},\theta_0,\ldots,\theta_{M-1})$ 
is obviously a property that is specific to Gibbs random field models.

Note that when we consider $M$ models from generic exponential families, this property of the concatenated sufficient
statistic rarely holds.
For instance, if under model $\mathcal{M}=0$, $x_i|\theta_0\overset{iid}{\sim} \mathcal{P}(\theta_0)$
and under model $\mathcal{M}=1$, $x_i| \theta_1 \overset{iid}{\sim} \mathcal{G}eo(\theta_1)$, this property is not satisfied 
since the distribution of $\bx$ given the common $S(\bx)=\sum_{i=1}^n x_i$ in the first model 
$$
h_0(\bx|S(\bx))=\left[ \sum_{\tilde\bx\in\mathcal{X}:S(\tilde\bx)=s}\frac{1}{\prod_{i=1}^n \tilde{x_i}!}\right]^{-1}\dfrac{1}{\prod_{i=1}^n x_i!}
$$
is different from  the distribution of $\bx$ given $S(\bx)$  in the other one
$$h_1(\bx|S(\bx))=\dfrac{1}{n(S(\bx))}.$$
As a consequence, $S(\bx)$ is not sufficient for the parameter $\mathcal{M}$.
For Gibbs random fields models, it is possible to apply the  ABC algorithm in order to produce an approximation with tolerance factor $\epsilon$:\\
\hrule
\smallskip
\noindent{\em ABC algorithm for model choice (ABC-MC):}
\begin{sffamily}
\begin{enumerate}
\item Generate  $m^{*}$ from the prior $\pi(\mathcal{M}=m)$.
 \item Generate $\theta_{m^{*}}^{*}$ from the prior $\pi_{m^{*}}(\cdot)$.
 \item Generate $\bx^{*}$ from the model $f_{m^*}(\cdot|\theta_{m^{*}}^{*})$.
 \item Compute the distance $\rho(S(\bx^0),S(\bx^{*}))$.
 \item Accept $(\theta_{m^{*}}^{*},m^{*})$ if $\rho(S(\bx^0),S(\bx^{*}))<\epsilon$, otherwise, start again in 1.
\end{enumerate}
\end{sffamily}
\hrule
\medskip

Simulating a data set $\bx^*$ from $f_{m^*}(\cdot|\theta_{m^{*}}^{*})$ at step 3 is often non-trivial for GRFs.
For the special case of the Ising model considered in the examples below, there have been many developments from \cite{Be74} to \cite{MoWa03}
that allow for exact simulation via perfect sampling. We refer the reader to \cite{Hagg02}, \cite{Mo03} and \cite{MoWa03},
for details of this simulation technique and for a discussion of its limitations.
For other GRFs it is often possible to use a Gibbs sampler updating one clique at a time conditional on the others.
This solution was implemented for the biophysical example of Section \ref{sec:application}.

For the same reason as above, this algorithm results in an approximate 
generation from the joint posterior distribution 
\begin{eqnarray*}
 \pi\left\{(\mathcal{M},\theta_0,\ldots,\theta_{M-1})|\rho(S(\bx^0),S(\bx^{*}))<\epsilon\right\}\,.
\end{eqnarray*}

When it is possible to achieve $\epsilon=0$, the algorithm is exact since $S$ is a sufficient
statistic. We have thus derived a likelihood-free method to handle model choice.

Once a sample of $N$ values of $(\theta_{m^{i*}}^{i*},m^{i*})$ $(1\le i\le N)$ is generated from this algorithm, a standard
Monte Carlo approximation of the posterior probabilities is provided by the empirical frequencies of visits
to the model, namely
$$
\widehat{\mathbb{P}}(\mathcal{M}=m|\bx^0)=\sharp\{m^{i*}=m\}\big/N\,,
$$
where $\sharp\{m^{i*}=m\}$ denotes the number of simulated $m^{i*}$'s equal to $m$.
Correlatively, the Bayes factor associated with the evidence provided by the data $\bx^0$ in favour of model $m_0$ 
relative to model $m_1$ is defined by
\begin{align}\label{eq:bf}
 &BF_{m_0/m_1}(\bx^0)=\frac{\mathbb{P}(\mathcal{M}=m_0|\bx^0)}{\mathbb{P}(\mathcal{M}=m_1|\bx^0)}
 \frac{\pi(\mathcal{M}=m_1)}{\pi(\mathcal{M}=m_0)}\\
 &=\frac{\int f_{m_0}(\bx^0|\theta_0)\pi_0(\theta_0)\pi(\mathcal{M}=m_0) 
\text{d}\theta_0}{\int f_{m_1}(\bx^0|\theta_1)\pi_1(\theta_1)\pi(\mathcal{M}=m_1) \text{d}\theta_1}
 \frac{\pi(\mathcal{M}=m_1)}{\pi(\mathcal{M}=m_0)}\,.
\end{align}
The previous estimates of the posterior probabilities can then be plugged-in to approximate the above Bayes factor by
\begin{eqnarray*}
\overline{BF}_{m_0/m_1}(\bx^0)&=&\frac{\hat{\mathbb{P}}(\mathcal{M}=m_0|\bx^0)}{\hat{\mathbb{P}}(\mathcal{M}=m_1|\bx^0)}\times
 \frac{\pi(\mathcal{M}=m_1)}{\pi(\mathcal{M}=m_0)}\\
&=&\dfrac{\sharp\{m^{i*}=m_0\}}{\sharp\{m^{i*}=m_1\}}\times\frac{\pi(\mathcal{M}=m_1)}{\pi(\mathcal{M}=m_0)}\,,
\end{eqnarray*}
but this estimate is only defined when $\sharp\{m^{i*}=m_1\}\ne 0$. To bypass this difficulty, the substitute
$$
 \widehat{BF}_{m_0/m_1}(\bx^0)=
\dfrac{1+\sharp\{m^{i*}=m_0\}}{1+\sharp\{m^{i*}=m_1\}}\times \frac{\pi(\mathcal{M}=m_1)}{\pi(\mathcal{M}=m_0)}
$$
is particularly interesting because we can evaluate its bias. (Note that there does not exist an unbiased estimator
of $BF_{m_0/m_1}(\bx^0)$ based on the $m^{i*}$'s.) Indeed, assuming without loss of generality that $\pi(\mathcal{M}=m_1)=
\pi(\mathcal{M}=m_0)$, if we set $N_0=\sharp\{m^{i*}=m_0\}$,  $N_1=\sharp\{m^{i*}=m_1\}$ then conditionally on $N=N_0+N_1$,
$N_1$ is a binomial $\mathcal{B}(N,p)$ rv with probability $p=(1+BF_{m_0/m_1}(\bx^0))^{-1}$. It is then straightforward
to establish that
$$
\mathbb{E} \left[\left. \dfrac{N_0+1}{N_1+1} \right| N \right] = BF_{m_0/m_1}(\bx^0)+\dfrac{1}{p(N+1)}-\dfrac{N+2}{p(N+1)}(1-p)^{N+1}\,.
$$
The bias in the estimator $\widehat{BF}_{m_0/m_1}(\bx^0)$ is thus $\{1-(N+2)(1-p)^{N+1}\}/(N+1)p$, which goes to zero as $N$ 
goes to infinity.

$\widehat{BF}_{m_0/m_1}(\bx^0)$ can be seen as the ratio of the posterior means on the model probabilities
$p$ under a $\mathcal{D}ir(1,\cdots,1)$ prior. In fact, if we denote $N_j=\sharp\{m^{i*}=m_j\}$, $N=\sum_{j=0}^{M-1}$ then the vector
$(N_1,\cdots,N_M)$ has a multinomial distribution 
$$
(N_0,\cdots,N_{M-1}|p_0,\cdots,p_{M-1})\sim \mathcal{M}(N;p_0,\cdots,p_{M-1})\,.
$$ 
The corresponding posterior distribution on $p$  is a $\mathcal{D}ir(1+N_0,\cdots,1+N_{M-1})$ and
$$
\widehat{BF}_{m_0/m_1}(\bx^0)=\dfrac{\mathbb{E}[p_0|N_0,\cdots,N_{M-1}]}{\mathbb{E}[p_1|N_0,\cdots,N_{M-1}]}=\dfrac{N_0+1}{N_1+1}
$$
is a consistent estimate of $BF_{m_0/m_1}(\bx^0)$.

Since the distribution of the sample $(\theta_{m^{i*}}^{i*},m^{i*})_{(1\le i\le N)}$ is not exactly
$\pi\left\{(\mathcal{M},\theta_0,\ldots,\allowbreak\theta_{M-1})|\bx^0\right\}$ but
$\pi\left\{(\mathcal{M},\theta_0,\ldots,\theta_{M-1})|\rho(S(\bx^0),S(\bx^{*}))<\epsilon\right\}$,
the Bayes factor should be written as
\begin{align*}
 &BF_{m_0/m_1}(\bx^0)=\frac{\mathbb{P}(\mathcal{M}=m_0|\rho(S(\bx^0),S(\bx^{*}))<\epsilon)}{\mathbb{P}(\mathcal{M}=m_1|\rho(S(\bx^0),S(\bx^{*}))<\epsilon)}
 \frac{\pi(\mathcal{M}=m_1)}{\pi(\mathcal{M}=m_0)}\\
&=\frac{\int \pi\left\{(\mathcal{M}=m_0,\theta_0)|\rho(S(\bx^0),S(\bx^{*}))<\epsilon\right\}\text{d}\theta_0}{\int\pi\left\{(\mathcal{M}=m_1,\theta_1)|\rho(S(\bx^0),S(\bx^{*}))<\epsilon\right\}\text{d}\theta_1}\frac{\pi(\mathcal{M}=m_1)}{\pi(\mathcal{M}=m_0)}
\\
&=\frac{\int \mathbb{P}_{\theta_0}(\rho(S(\bx^0),S(\bx^{*}))<\epsilon)\pi_0(\theta_0)\pi(\mathcal{M}=m_0)\text{d}\theta_0}
{\int \mathbb{P}_{\theta_1}(\rho(S(\bx^0),S(\bx^{*}))<\epsilon)\pi_1(\theta_1)\pi(\mathcal{M}=m_1)\text{d}\theta_1}
\frac{\pi(\mathcal{M}=m_1)}{\pi(\mathcal{M}=m_0)}
\\
&=\frac{\int \left[ \int f_{m_0}(\bx^*|\theta_0)\pi_0(\theta_0) \text{d}\theta_0 \right]\mathbb{I}_{\left\lbrace \rho(S(\bx^0),S(\bx^{*}))<\epsilon\right\rbrace } \text{d}\bx^*}{\int \left[ \int f_{m_1}(\bx^*|\theta_1)\pi_1(\theta_1) \text{d}\theta_1 \right]\mathbb{I}_{\left\lbrace \rho(S(\bx^0),S(\bx^{*}))<\epsilon\right\rbrace } \text{d}\bx^*}
\end{align*}
When  $\epsilon=0$ and $S(\bx)$ is a sufficient statistic, this expression corresponds to equation (\ref{eq:bf}). 

\subsection{Two step ABC}
The above estimator $ \widehat{BF}_{m_0/m_1}(\bx^0)$ is rather unstable (i.e.~it suffers from a large variance)
when $BF_{m_0/m_1}(\bx^0)$ is very large since, when 
$\mathbb{P}(\mathcal{M}=m_1|\bx^0)$ is very small, $\sharp\{m^{i*}=m_1\}$ is most often equal to zero. 
This difficulty can be bypassed by a reweighting scheme. If the choice of $m^{*}$ in the ABC algorithm is
driven by the probability distribution $\mathbb{P}(\mathcal{M}=m_1)=\varrho=1-\mathbb{P}(\mathcal{M}=m_0)$ 
rather than by $\pi(\mathcal{M}=m_1)=1-\pi(\mathcal{M}=m_0)$, the value of $\sharp\{m^{i*}=m_1\}$ can be
increased and later corrected by considering instead
$$
\widetilde{BF}_{m_0/m_1}(\bx^0)=
\dfrac{1+\sharp\{m^{i*}=m_0\}}{1+\sharp\{m^{i*}=m_1\}}\times \frac{\varrho}{1-\varrho}\,.
$$
Therefore, if a first run of the ABC algorithm exhibits a very large value of $ \widehat{BF}_{m_0/m_1}(\bx^0)$, the
estimate $\widetilde{BF}_{m_0/m_1}(\bx^0)$ produced by a second run with 
$$
\varrho \propto 1\bigg/\hat{\mathbb{P}}(\mathcal{M}=m_1|\bx^0)
$$
will be more stable than the original $\widehat{BF}_{m_0/m_1}(\bx^0)$. In the most extreme cases when no $m^{i*}$ is ever
equal to $m_1$, this corrective second is unlikely to bring much stabilisation, though.
>From a practical point of view, obtaining a poor evaluation of $BF_{m_0/m_1}(\bx^0)$ when the Bayes factor is very small (or very large) has
limited consequences since the poor approximation also leads to the same conclusion about the choice of model $m_0$. 
Note, however, that, when there are more than two models, using these approximations to perform Bayesian model averaging can be dangerous.

\section{Results}
\label{sec:res}
\subsection{Toy example}
Our first example compares an iid Bernoulli model with a two-state first-order Markov chain. Both models are special cases
of GRF, the first one with a trivial neighbourhood structure and the other one with a  nearest neighbourhood structure.
Furthermore, the normalising constant $Z_{\theta_m,m}$ can be computed in closed form, as well as the
posterior probabilities of both models. We thus consider a sequence $\mathbf{x}=(x_1,..,x_n)$ of binary variables. Under 
model $\mathcal{M}=0$, the GRF representation of the Bernoulli distribution $\mathcal{B}(\exp(\theta_0)/\{1+\exp(\theta_0)\})$ is
$$
f_0(\mathbf{x}|\theta_0)=\exp\left(\theta_0\sum_{i=1}^{n}\mathbb{I}_{\{x_i=1\}} \right)\bigg/ \{1+\exp(\theta_0)\}^n\,,
$$
associated with the sufficient statistic $S_0(\mathbf{x})=\sum_{i=1}^{n}\mathbb{I}_{\{x_i=1\}}$ and the normalising constant
$Z_{\theta_0,0}=(1+e^{\theta_0})^n$. Under a uniform prior $\theta_0\sim\mathcal{U}(-5,5)$, the posterior probability of this model is
available since the marginal when $S_0(\mathbf{x})=s_0$ $(s_0\ne 0)$ is given
by
$$
\dfrac{1}{10}\, \sum_{k=0}^{s_0-1} {s_0-1\choose k} \frac{(-1)^{s_0-1-k}}{n-1-k}
\left[ (1+e^5)^{k-n+1}-(1+e^{-5})^{k-n+1} \right]\,,
$$
by a straightforward rational fraction integration.

Model $\mathcal{M}=1$ is chosen as a Markov chain (hence a particular GRF in dimension one with 
$i$ and $i'$ being neighbours if $|i-i'|=1$) with the special feature that the probability 
to remain within the same state is constant over both states, namely 
$$
\mathbb{P}(x_{i+1}=x_i|x_i)=\exp(\theta_1)\big/\{1+\exp(\theta_1)\}\,.
$$ 
We assume a uniform distribution on $x_1$ and the likelihood function for this model is thus
$$
f_1(\mathbf{x}|\theta_1) =\frac{1}{2}\exp\left(\theta_1 \sum_{i=2}^{n}\mathbb{I}_{\{x_i=x_{i-1}\}}\right)
\bigg/\{1+\exp(\theta_1)\}^{n-1}\,,
$$
with $S_1(\mathbf{x})=\sum_{i=2}^{n}\mathbb{I}_{\{x_i=x_{i-1}\}}$ being the sufficient statistic and $Z_{\theta_1,1}=2(1+e^{\theta_0})^{n-1}$
being the normalising constant in that case. Under a uniform prior $\theta_1\sim\mathcal{U}(0,6)$, the posterior probability of this model is
once again available, the likelihood being of the same form as when $\mathcal{M}=0$. The bounds of the prior distributions on $\theta_0$ and $\theta_1$
were chosen to avoid data sets consisting in a sequence of $n$ identical values since it is impossible to distinguish model $0$ and model $1$ in that case.

\begin{figure}
\centerline{\includegraphics[height=7cm, width=13cm]{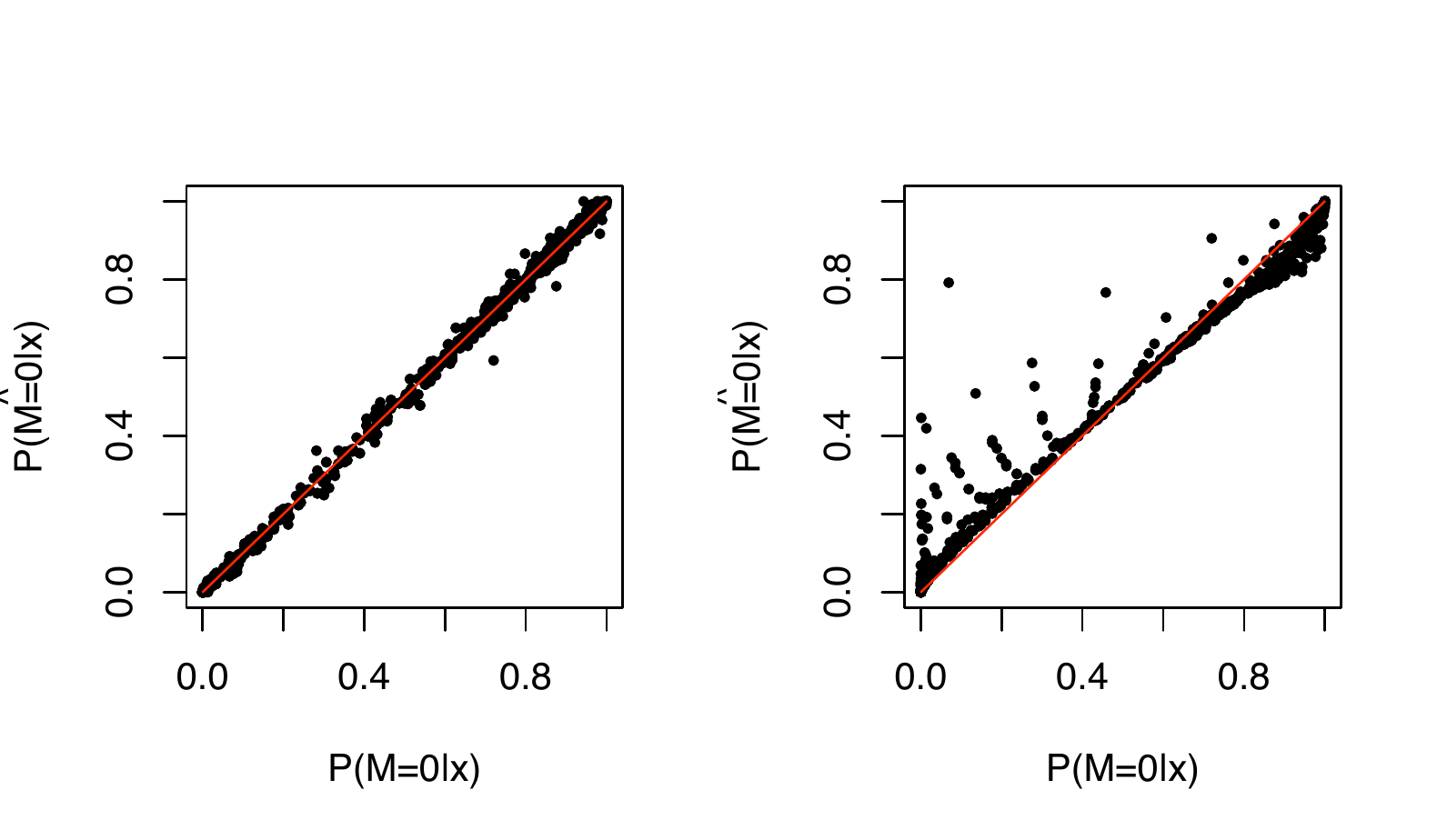}}
\caption{\label{fig:prob}
{\textit{(left)}} Comparison of the true $\mathbb{P}(\mathcal{M}=0|\bx^0)$ with $\widehat{\mathbb{P}}(\mathcal{M}=0|\bx^0)$  over $2,000$
simulated sequences and $4\times 10^6$ proposals from the prior. The red line is the diagonal. {\textit{(right)}} Same comparison when using
a tolerance $\epsilon$ corresponding to the $1\%$ quantile on the distances.}
\end{figure}

\begin{figure}
\centerline{\includegraphics[height=7cm, width=13cm]{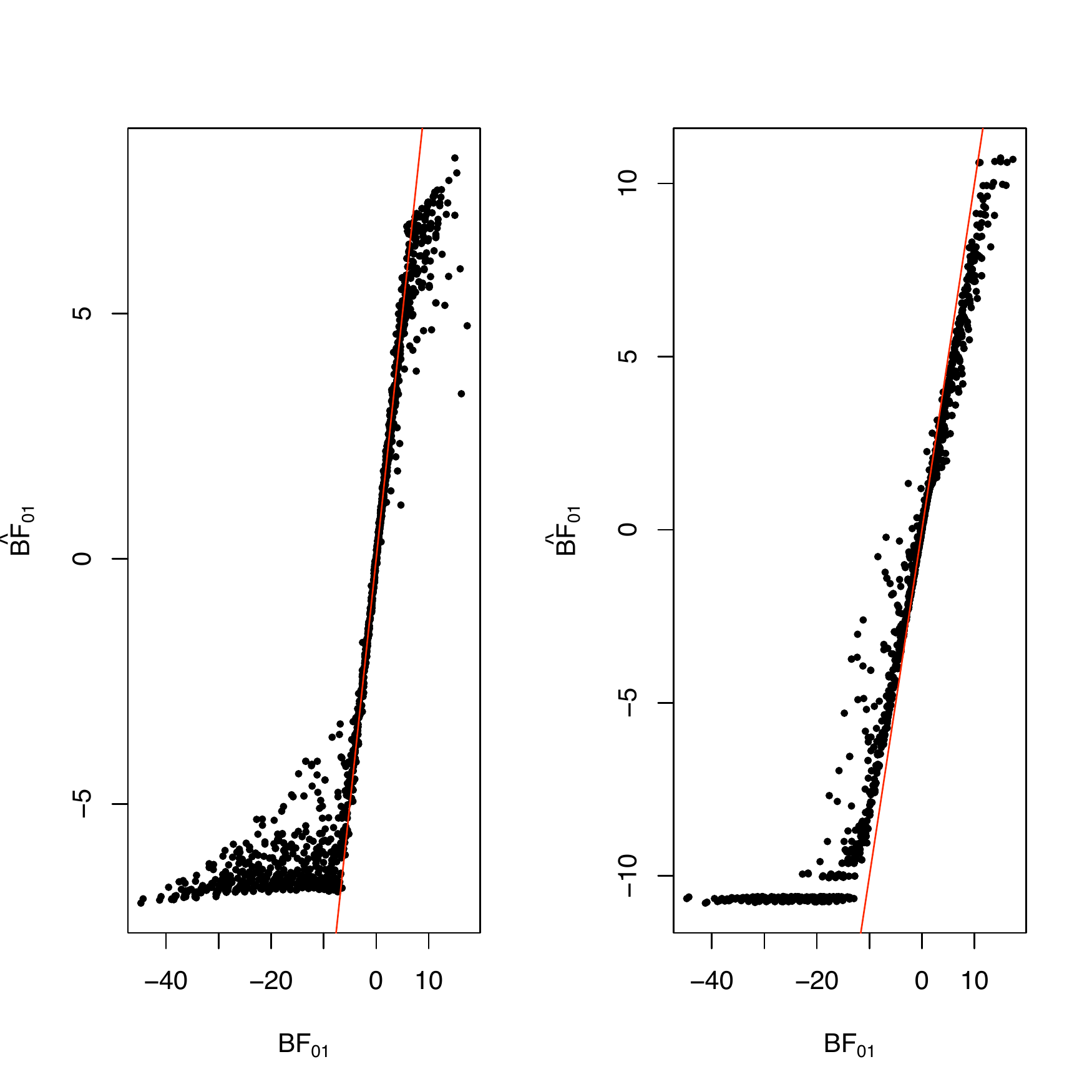}}
\caption{\label{fig:BF}
{ \textit{(left)}} Comparison of the true ${BF}_{m_0/m_1}(\bx^0)$ with $\widehat{BF}_{m_0/m_1}(\bx^0)$ (in logarithmic scales) over $2,000$
simulated sequences and $4\times 10^6$ proposals from the prior. The red line is the diagonal. { \textit{(right)}} Same comparison when using
a tolerance corresponding to the $1\%$ quantile on the distances.}
\end{figure}

\begin{figure}
\centerline{\includegraphics[height=7.5truecm]{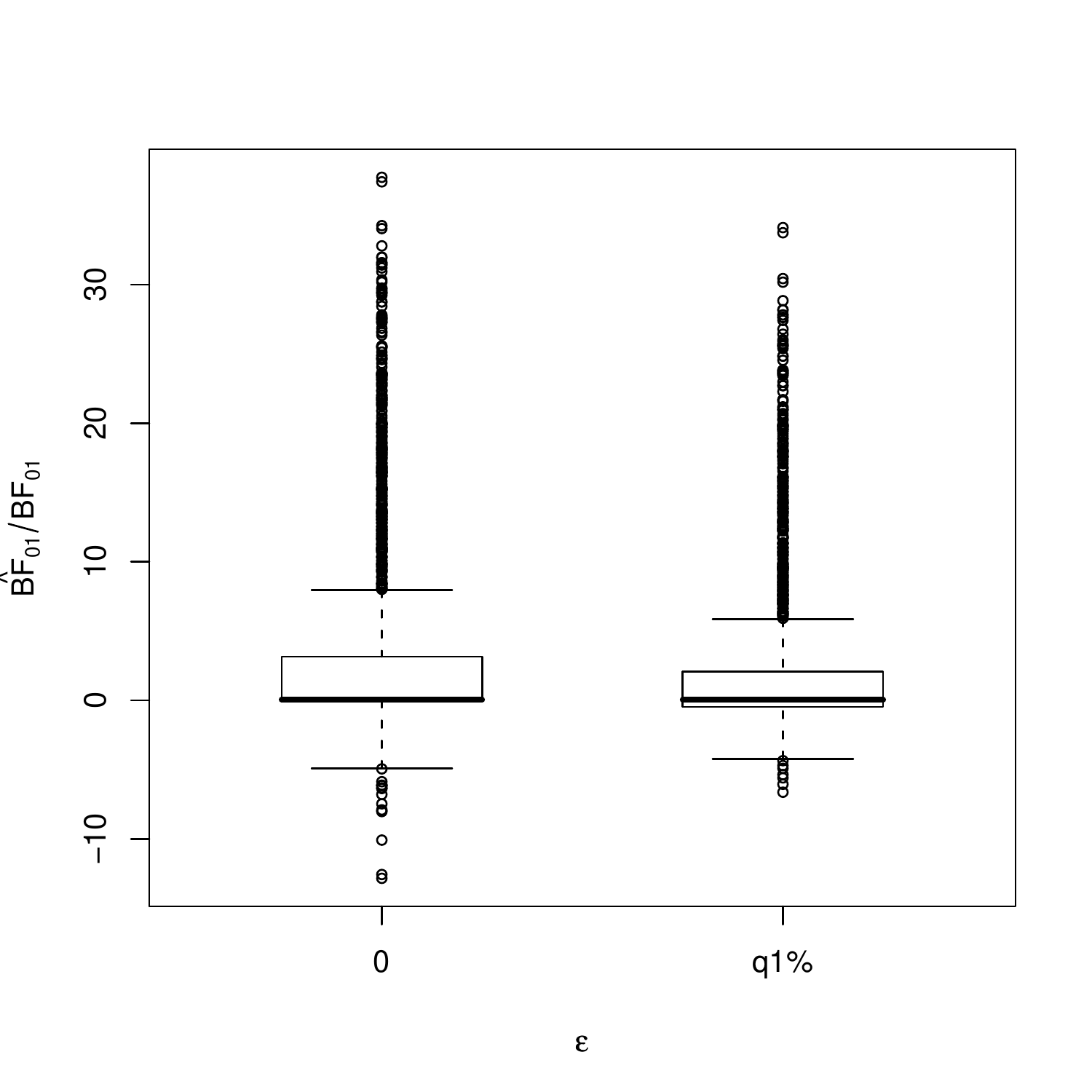}}
\caption{\label{fig:boxplot}{ \textit{(left)}} Boxplots of  the ratios $\widehat{BF}_{m_0/m_1}(\bx^0)/{BF}_{m_0/m_1}(\bx^0)$ (in logarithmic scales)
in the exact case and using a tolerance equal to the $1\%$ quantile on the distances over $2,000$
simulated sequences and $4\times 10^6$ proposals from the prior.}
\end{figure}

We are therefore in a position to evaluate the ABC approximations of the model posterior probabilities and of the Bayes factor against
the exact values. For this purpose, we simulated $1000$ datasets $\mathbf{x}^0=(x_1,\cdots,x_n)$ with $n=100$ under each model, using parameters
simulated from the priors and computed the exact posterior probabilities and the Bayes factors in both cases. 
For each of those $2000$ datasets $\mathbf{x}^0$, the ABC-MC algorithm was run for
$4\times 10^6$ loops, meaning that $4\times 10^6$ sets $(m^*,\theta^*_{m^*},\mathbf{x}^*)$ were exactly simulated from the joint distribution and a random number of those were accepted
when $S(\mathbf{x}^*)=S(\mathbf{x}^0)$. (In the worst case scenario, the number of acceptances was 12!)
As shown  on the left graph of Figure \ref{fig:prob}, the fit of the approximate posterior probabilities is good for all values of $\mathbb{P}(\mathcal{M}=0|\bx^0)$.
When we introduce a tolerance $\epsilon$ equal to the $1\%$ quantile of  $\rho(S(\bx^0),S(\bx^{*}))$, $\rho$ being the Euclidean distance,
the results are similar when $\mathbb{P}(\mathcal{M}=0|\bx^0)$ is close to $0$, $1$ or $0.5$, and we observe a slight 
difference for other values.  We also evaluated the approximation of the Bayes factor (and of the subsequent model choice) 
against the exact Bayes factor. As clearly pictured on the left graph of Figure \ref{fig:BF}, the fit is good in the exact
case ($\epsilon=0$), the poorest fits occurring in the limiting cases when the Bayes factor is either very 
large or very small and thus when the model choice is not an issue, as noted above. In the central zone when $\log {BF}_{m_0/m_1}(\bx^0)$ is close
to $0$, the difference is indeed quite small, the few diverging cases being due to occurrences of very small acceptance rates. If we classify the values of  ${BF}_{m_0/m_1}(\bx^0)$ and $\widehat{BF}_{m_0/m_1}(\bx^0)$ according to the Jeffrey's scale, we observe that the Bayes factor and its approximation belong to the same category ($1903$ simulated data sets are on the diagonal of Table \ref{tab:bf_epsilon0}) or to very close categories. Once more, using a tolerance $\epsilon$ equal to the $1\%$ quantile does not bring much difference in the output, Table \ref{tab:bf_epsilon1} shows that the Bayes factor and its estimation still belong to the same category for $1805$ simulated data sets. The approximative Bayes factor is slightly less discriminative in that case, since the slope of the cloud is less than
the unitary slope of the diagonal on the right graph of Figure \ref{fig:BF}; ${BF}_{m_0/m_1}(\bx^0)$ and $\widehat{BF}_{m_0/m_1}(\bx^0)$ lead to the selection of the same model, but with a lower degree of confidence for the second one (Table \ref{tab:bf_epsilon1}). 
The boxplots on Figure \ref{fig:boxplot} compare the distributions of the ratios $\widehat{BF}_{m_0/m_1}(\bx^0)/{BF}_{m_0/m_1}(\bx^0)$ in the exact case and using a tolerance equal to the $1\%$ quantile on the distances. As reported in Table \ref{tab:quantiles}, the median is very close to $1$ in both cases. The ratio takes more often extreme values in the exact case. Once more, this is a consequence of the poor estimation of the Bayes factor when the acceptance rate is small. 
Given that using the tolerance version allows for more simulations to be used in
the Bayes factor approximation, we thus recommend using this approach.

\begin{table}
\begin{center}
\begin{tabular}{c|c|c|c|c|c|c|c|c}
&$m=1$&$m=1$&$m=1$&$m=1$&$m=0$&$m=0$&$m=0$&$m=0$\\ 
&dec.&str.&sub.&weak&weak&sub.&str.&dec.\\ \hline
$m=1$, dec.&778&9&0&0&0&0&0&0\\ \hline
$m=1$, str.&2&79&0&0&0&0&0&0\\ \hline
$m=1$, sub.&0&7&53&0&0&0&0&0\\ \hline
$m=1$, weak&0&0&2&63&0&7&0&0\\ \hline
$m=0$, weak&0&0&0&22&103&7&0&0\\ \hline
$m=0$, sub.&0&0&0&0&1&103&23&0\\ \hline
$m=0$, str.&0&0&0&0&0&5&177&6\\ \hline
$m=0$, dec.&0&0&0&0&0&0&13&547\\ \hline
\end{tabular}
\end{center}
\caption{\label{tab:bf_epsilon0} Comparison of the decisions based on ${BF}_{m_0/m_1}(\bx^0)$ and on
$\widehat{BF}_{m_0/m_1}(\bx^0)$ using $\epsilon=0$ according to the Jeffrey's scale
(dec.: decisive $\log({BF}_{m_0/m_1}(\bx^0))>2$, str.: strong $1<\log({BF}_{m_0/m_1}(\bx^0))<2$,
sub.: substantial $0.5<\log({BF}_{m_0/m_1}(\bx^0))<1$, weak $0<\log({BF}_{m_0/m_1}(\bx^0))<0.5$).}
\end{table}

\begin{table}
\begin{center}
\begin{tabular}{c|c|c|c|c|c|c|c|c}
&$m=1$&$m=1$&$m=1$&$m=1$&$m=0$&$m=0$&$m=0$&$m=0$\\ 
&dec.&str.&sub.&weak&weak&sub.&str.&dec.\\ \hline
$m=1$, dec.&740&39&5&2&0&0&1&0\\ \hline
$m=1$, str.&0&64&14&2&1&0&0&0\\ \hline
$m=1$, sub.&0&0&39&19&2&0&0&0\\ \hline
$m=1$, weak&0&0&0&61&3&0&1&0\\ \hline
$m=0$, weak&0&0&0&2&128&2&0&0\\ \hline
$m=0$, sub.&0&0&0&0&2&123&1&1\\ \hline
$m=0$, str.&0&0&0&0&0&26&161&1\\ \hline
$m=0$, dec.&0&0&0&0&0&0&71&489\\ \hline
\end{tabular}
\end{center}
\caption{\label{tab:bf_epsilon1} Comparison of the decisions based on ${BF}_{m_0/m_1}(\bx^0)$ and on $\widehat{BF}_{m_0/m_1}(\bx^0)$
according to the Jeffrey's scale, using a tolerance $\epsilon$ corresponding to the $1\%$ quantile of the distances
(dec.: decisive $\log({BF}_{m_0/m_1}(\bx^0))>2$, str.: strong $1<\log({BF}_{m_0/m_1}(\bx^0))<2$,
sub.: substantial $0.5<\log({BF}_{m_0/m_1}(\bx^0))<1$, weak $0<\log({BF}_{m_0/m_1}(\bx^0))<0.5$).}
\end{table}

\begin{table}
 \begin{center}
\begin{tabular}{c|c|c|c}
&$q_{0.25}$&$q_{0.5}$&$q_{0.75}$ \\ \hline
$\epsilon=0$&$0.914$&$1.041$&$22.9$\\ \hline
$\epsilon=q_{1\%}$&$0.626$&$1.029$&$7.9$
\end{tabular}
\end{center}
\caption{\label{tab:quantiles} Quantiles of the ratios $\widehat{BF}_{m_0/m_1}(\bx^0)/{BF}_{m_0/m_1}(\bx^0)$ in the exact
case and using a tolerance $\epsilon$ equal to the $1\%$ quantile of the distances.}
\end{table}

\subsection{Application to protein 3D structure prediction}
\label{sec:application}

The numerous genome sequences now available provide a huge amount of
protein sequences whose functions remain unknown. A classical strategy is to determine
the tridimensional (3D) structure of a protein, also called {\em fold,} as it provides important and valuable information about its function.
Experimental methods, like those based on X-ray diffraction or nuclear magnetic resonance, provide accurate
descriptions of 3D-structures, but are time consuming. As an alternative, computational methods have been proposed to predict 3D structures.

These latter methods mostly rely on homology (two proteins are said
to be homologous if they share a common ancestor). In fact,  homologous proteins often share similar function and, as function is controlled by structure, similar structure.
When the protein under study, hereafter called the query protein,
can be considered as homologous with another protein, a prediction of its 3D structure based on the structure of its homolog can be built.

First, one compares the sequence of the query protein with a data bank of sequences of proteins of known structures but
sequence similarity is often too low to assess homology with sufficient certainty. Because of selection pressure on the function,
structures are more conserved over time than sequences. {\em Threading} methods consist in aligning the query sequence onto a set
of structures representative of all known folds. The sequence of the query is threaded onto the candidate structures in order to find
the most compatible one. A score (a fitting criterion) is computed for each proposal. Structures displaying sufficiently high scores,
if any, are chosen as the corresponding protein can be said homologous with the query protein.

It may happen that both information based on sequence similarity and threading score are not sufficient to access protein homology
and consequently, to select a  3D structure. Our aim is to use extra information to help making a decision, if necessary. We use here
the fact that amino acids in close contact in the 3D structure often share similar (or complementary) biochemical properties.
In the example we discuss in this section, we use hydrophobicity as a clustering factor since hydrophobic amino-acids are mostly
buried inside the 3D structure, and hydrophilic ones exposed to water. This effect is observed in almost all proteins.

>From a formal perspective, each structure can be represented by a
graph where a node represents one amino-acid of the protein and an
edge between two nodes indicates that both amino-acids are in close
contact in the folded protein (hence are neighbours). Labels are allocated to each node, associated with
hydrophobicity of amino-acids (amino-acids are classified as
hydrophobic or hydrophilic according to Table \ref{hydrophobicity}). Then, a Gibbs random field, more
precisely an Ising model, can be defined on each graph.
When several structures are proposed by a threading method, the ABC-MC
algorithm is then available to select the most likely structure.

\begin{table}
\caption{\label{hydrophobicity} Classification of amino acids into hydrophilic and hydrophobic groups.}

\begin{center}
\begin{tabular}{c|c}
Hydrophilic & Hydrophobic \\ \hline
K E R D Q N P H S T G&A Y M W F V L I C
\end{tabular}
\end{center}
\end{table}

We applied this procedure to proteins of known structures (here called the native structures) \textbf{1tqgA}, involved into signal transduction
processes in the bacterium \textit{Thermotoga maritima} and \textbf{1k77A} which is a putative oxygenase from \textit{Escherichia coli}.
In these studies, the sequences were treated as queries, since our purpose was to evaluate if our idea could help in real 
situations.

We used FROST~\citep{Ma02}, a software dedicated to threading, and
MODELLER~\citep{MODELLER} to find the candidate structures and KAKSI~\citep{Mart05} to build the graphs. All candidate structures were picked up from the
Protein Data Bank ({\sf http://www.rcsb.org/pdb/home/home.do}). FROST provides the best alignment of the query sequence onto a structure, based on
score optimisation, and  the final score  measures  alignment quality. A score larger than $9$ means that the alignment is good,
while a score less than $7$ means the opposite.  For values between $7$ and $9$, this score cannot be used to reach a decision.
Additionally, FROST calculates the percentages of identity between query and candidate sequences; sequences with a percentage of sequence identity
higher than $>20\%$ can be considered as homologous.

As the native structures were known, similarities between candidate and native structures could be assessed, here by the
TM-score, \citep{TM}. A score larger than $0.4$
implies both structures are similar and a score
less than $0.17$ means that the prediction is nothing more than a
random selection from the PDB library.

\begin{figure}
\centerline{\includegraphics[height=5cm]{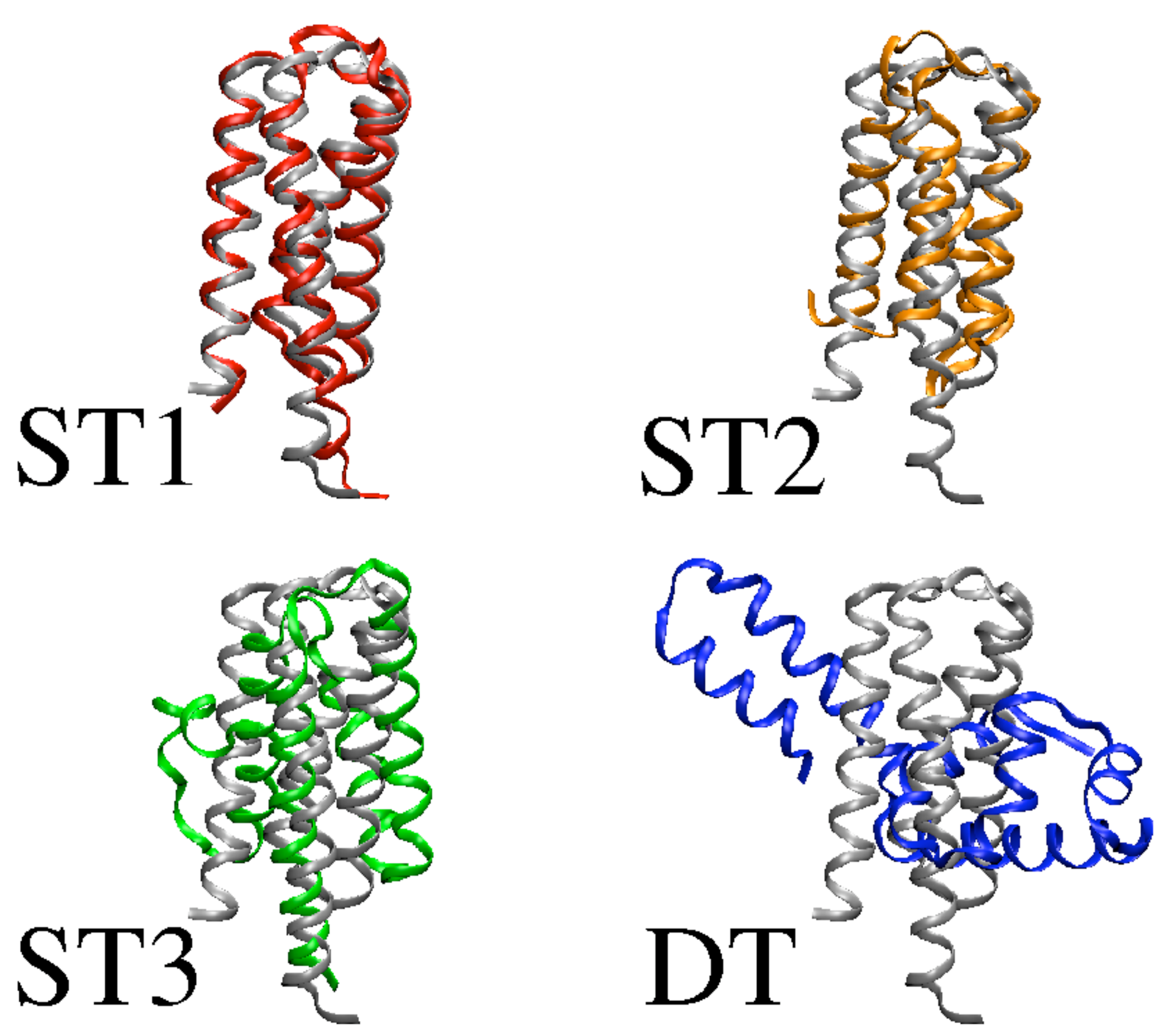}}
\vglue .4truecm
\caption{\label{fig:superPo} Superposition of the native structure
of \textbf{1tqgA} {\em (grey)} with the \textbf{ST1} structure {\em
(red)}, the \textbf{ST2} structure {\em (orange)}, the \textbf{ST3}
structure {\em (green)}, and the {\bf DT} structure {\em (blue)}.}
\end{figure}

\begin{figure}
\centerline{\begin{tabular}{c c}
 \includegraphics[height=2.5cm]{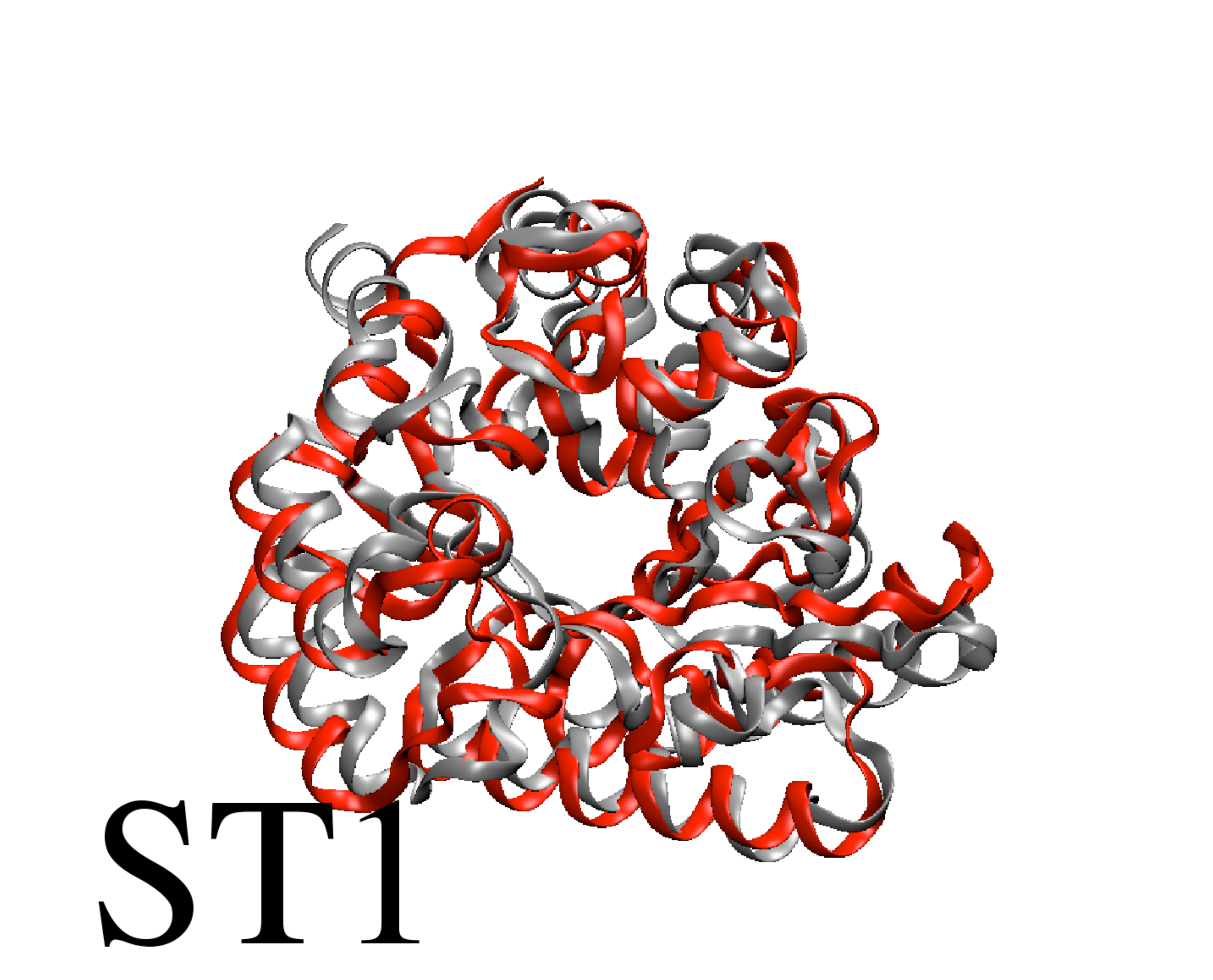} &\includegraphics[height=2.5cm]{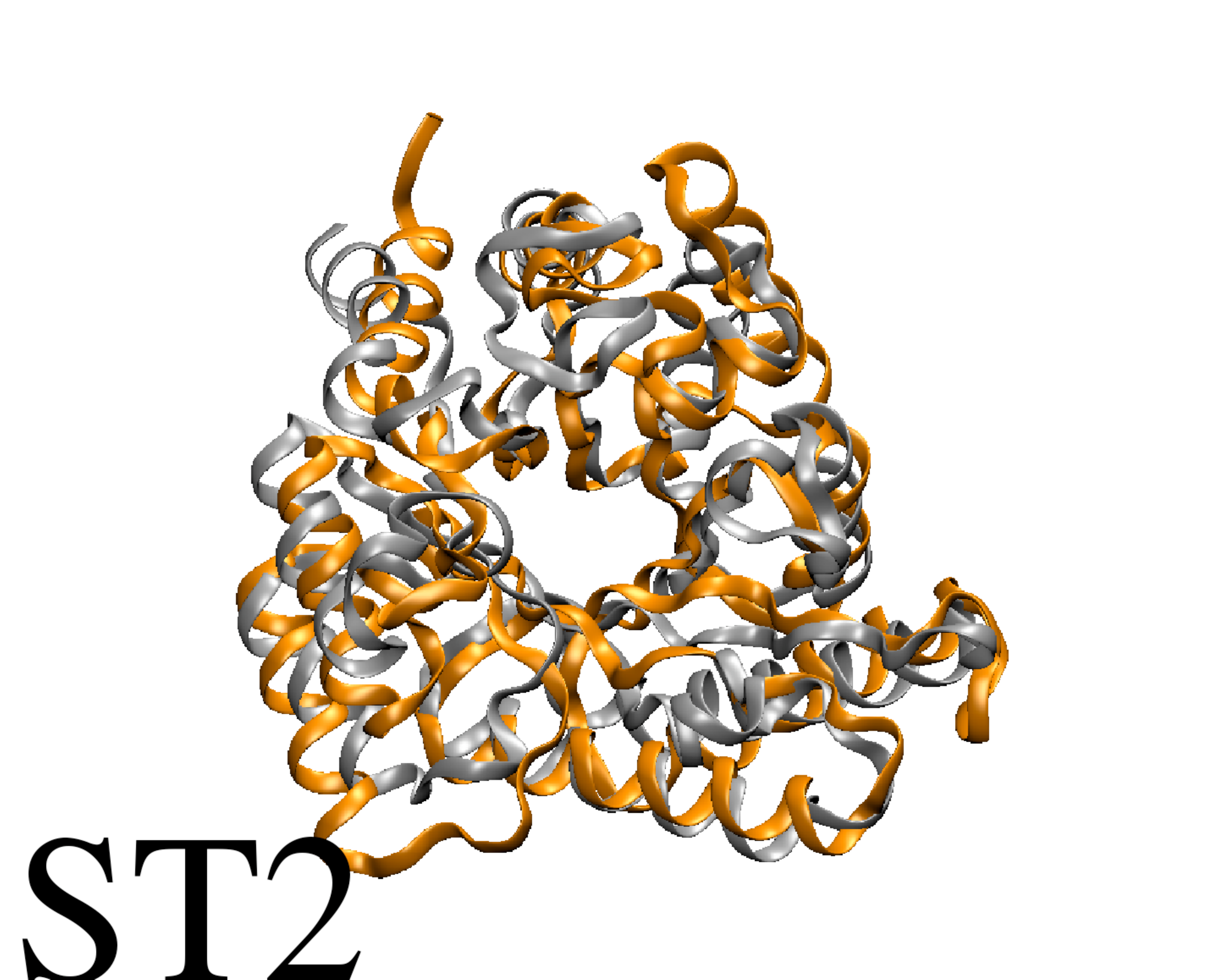}\\ \includegraphics[height=2.5cm]{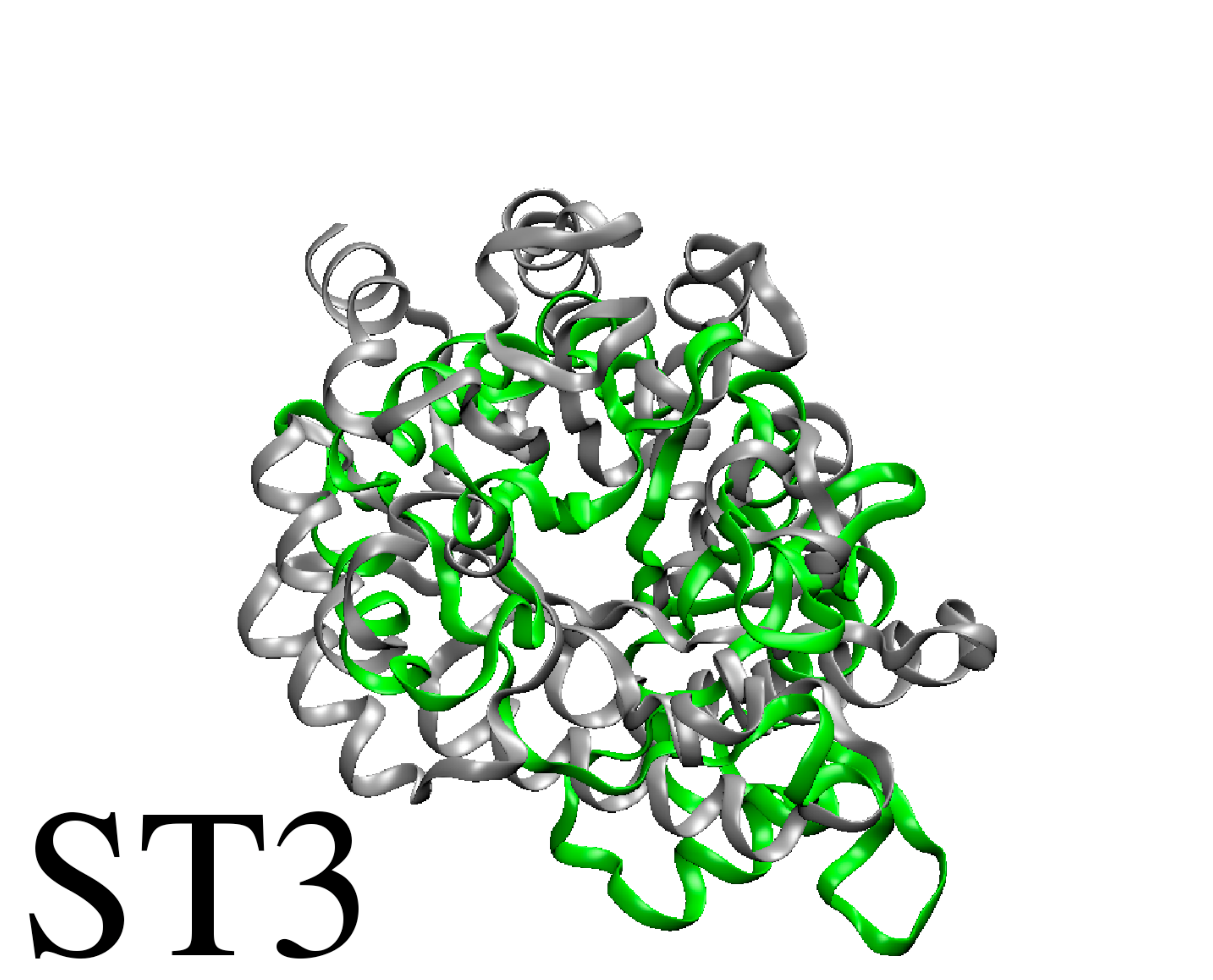} &\includegraphics[height=2.5cm]{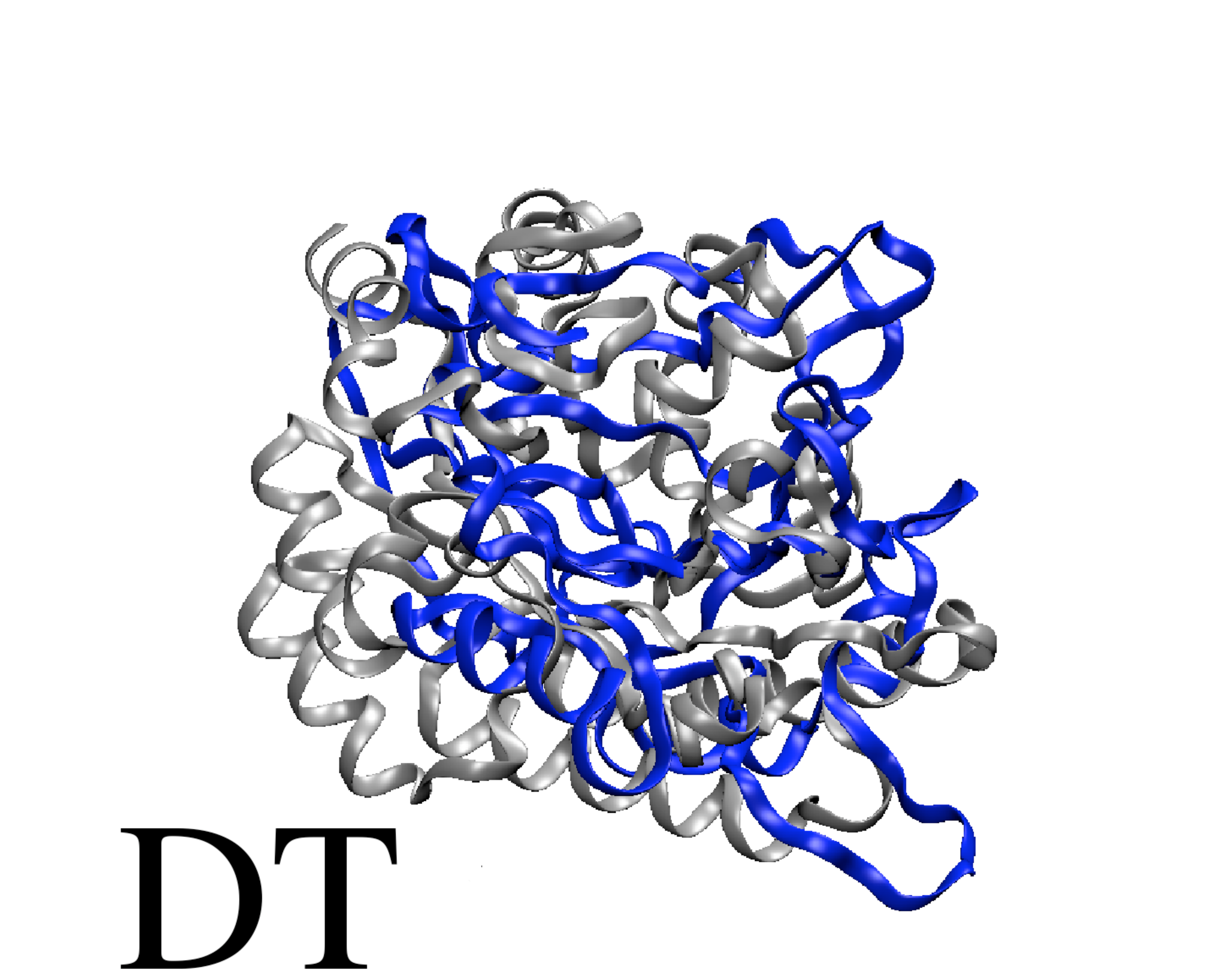}
 \end{tabular}
}
\vglue .4truecm
\caption{\label{fig:superPo2} Superposition of the native structure of \textbf{1k77A}
{\em (grey)} with the \textbf{ST1} structure {\em (red)}, the
\textbf{ST2} structure {\em (orange)}, the \textbf{ST3} structure {\em
(green)}, and the {\bf DT} structure {\em (blue)}.}
\end{figure}

For each query protein, we selected four candidates, called \textbf{ST1}, \textbf{ST2}, \textbf{ST3} and \textbf{DT},  covering
the whole spectrum of predictions that can be generated by protein
threading, from good to very poor \citep{Ta08} as described in Table
\ref{dataset} and \ref{dataset2}.
We selected essentially candidate structures for which no decision could have been made since they were scored in the FROST uncertainty zone.
According to the TM-score, \textbf{ST1} and \textbf{ST2} are considered as similar to the native structure, while \textbf{ST3} and \textbf{DT} are not.
For \textbf{ST1} and \textbf{ST2},
the alignment of the query sequence onto the candidate structure is good or fair; sequence  similarity  is higher for \textbf{ST1} than \textbf{ST2}. 
\textbf{ST3} is a poorer candidate since it is certainly not an homolog of the
query and the alignment is much poorer. For \textbf{DT}, the query sequence has been aligned with a structure that only shares few
structural elements with the native structure. Differences between the native structures and the corresponding
predicted structures are illustrated on Figure \ref{fig:superPo} for \textbf{1tqgA} and on Figure  \ref{fig:superPo2} for \textbf{1k77A}.

\begin{table}
\begin{center}
\begin{tabular}{l|c|c|c}
  &\% seq. Id.  & TM-score &FROST score  \\ \hline
 1i5nA (\textbf{ST1})& $32$ &$0.86$& $75.3$ \\ \hline
1ls1A1 (\textbf{ST2}) & $5$ &$0.42$ & $8.9$\\ \hline
 1jr8A (\textbf{ST3})&  $4$&$0.24$&  $8.9$\\ \hline
 1s7oA (\textbf{DT})&  $10$& $0.08$&$7.8$
\end{tabular}
\end{center}
\caption{\label{dataset} Summary of the characteristics of our
dataset for the protein \textbf{1tqgA}. {\em \% seq. Id.}: percentage of identity with the query
sequence. {\em TM-score}: similarity between a predicted structure
and the native structure. {\em FROST score}: quality of the alignment
of the query onto the candidate structure.}
\end{table}

\begin{table}
\begin{center}
\begin{tabular}{l|c|c|c}
  &\% seq. Id.  & TM-score &FROST score  \\ \hline
 1i60A (\textbf{ST1})& $16$ &$0.69$& $8.9$ \\ \hline
1qtwA (\textbf{ST2}) & $6$ &$0.54$ & $9.8$\\ \hline
 1qpoA1 (\textbf{ST3})&  $9$&$0.29$&  $9.3$\\ \hline
 1m4oA (\textbf{DT})&  $7$& $0.17$&$8.3$
\end{tabular}
\end{center}
\caption{\label{dataset2} Summary of the characteristics of our
dataset for the protein \textbf{1k77A}.  {\em \% seq. Id.}: percentage of identity with the query
sequence.  {\em TM-score}: similarity between a predicted structure
and the native structure. {\em FROST score}: quality of the alignment
of the query onto the candidate structure.}
\end{table}

Using ABC-MC, we then estimate the Bayes factors between model
\textbf{NS} corresponding to the true structure and models
\textbf{ST1}, \textbf{ST2}, \textbf{ST3}, and \textbf{DT},
corresponding to the predicted structures.
Each model is an Ising model with sufficient statistic
$S_m(\bx)=\sum_{i\overset{m}{\sim} i'} \mathbb{I}_{\{x_i=x_{i'}\}}$.
The scalar parameter $\theta_m$ of the Ising model $m$ is assumed to have a uniform prior
on the interval $[0,4]$. Simulated data sets were obtained by a standard Gibbs sampler.
The Gibbs algorithm has been iterated 1000 times, which is a sufficient
number of iterations for stabilisation.
We picked $\epsilon$ as the empirical $1\%$\_quantile of
the Euclidean distance $\rho(S(\bx^0),S(\bx^*))$.

Estimated values for the Bayes factors of model \textbf{NS} against each alternative are
given in Tables \ref{resdataset} and \ref{resdataset2}. As expected, all estimated Bayes factors are
larger than $1$  indicating that the data is always
in favour of the native structure, when compared with one of the four
alternatives and  Bayes factors increase when the similarity between candidate and native structure is lower.
Moreover, we can classify  the candidate structures into two categories: for \textbf{ST1} and \textbf{ST2},
the evidence is weak in favour of the native structure while the evidence is substantial or strong when the
alternative is \textbf{ST3} or \textbf{DT}. Thus our approach can distinguish similar from dissimilar structures,
even when they were scored in the FROST uncertainty zone.

\begin{table}
\begin{center}
\begin{tabular}{c|c|c|c|c}
&\textbf{NS}/\textbf{ST1} &\textbf{NS}/\textbf{ST2} &\textbf{NS}/\textbf{ST3} &\textbf{NS}/\textbf{DT}\\ \hline
$\widehat{BF}$&$1.34$ &$1.22$ &$2.42$ &$2.76$
\end{tabular}
\end{center}
\caption{\label{resdataset} Estimates of the Bayes factors between
model \textbf{NS} and models \textbf{ST1}, \textbf{ST2}, \textbf{ST3},
and \textbf{DT}, 
based on an ABC-MC algorithm using $1,2\times 10^6$ simulations
and a tolerance $\epsilon$ equal to the $1\%$ quantile of the
distances for the query protein \textbf{1tqgA}.}
\end{table}

\begin{table}
\begin{center}
\begin{tabular}{c|c|c|c|c}
&\textbf{NS}/\textbf{ST1} &\textbf{NS}/\textbf{ST2} &\textbf{NS}/\textbf{ST3} &\textbf{NS}/\textbf{DT}\\ \hline
$\widehat{BF}$&$1.07$ &$1.14$ &$11997$ &$14.24$
\end{tabular}
\end{center}
\caption{\label{resdataset2} Estimates of the Bayes factors between
model \textbf{NS} and models \textbf{ST1}, \textbf{ST2}, \textbf{ST3},
and \textbf{DT}, 
\textbf{NS} based on an ABC-MC algorithm using $1,2\times 10^6$ simulations
and a tolerance $\epsilon$ equal to the $1\%$ quantile of the
distances for the query protein \textbf{1k77A}.}
\end{table}

\section{Discussion}
This paper has hopefully demonstrated that the auxiliary variable technique that supports the ABC algorithm can
be used to overcome the lack of closed-form normalising constants in Gibbs random field models and in particular 
in Ising models. The computation of Bayes factors can therefore follow from a standard Monte Carlo simulation 
that includes the model index without requiring advanced techniques like reversible jump moves \citep{RC04}.
The usual approximation inherent to ABC methods can furthermore be avoided due to the availability of a sufficient statistic
across models. However, the toy example studied above shows that the accuracy of the approximation to the posterior probabilities 
and to the Bayes factor can be greatly improved by resorting to the original ABC approach, since it allows for the inclusion of many
more simulations. In the biophysical application to the choice of a folding structure for two proteins, we have also demonstrated that
we can implement the ABC solution on realistic datasets and, in the examples processed there, that the Bayes factors allow for
a ranking more standard methods do not.
 
\section*{Acknowledgements}
Aude Grelaud is a PhD candidate at Universit\'e Paris Dauphine.
The authors' research is partly supported by the Agence Nationale de la Recherche (ANR, 212,
rue de Bercy 75012 Paris) through the 2005 project ANR-05-BLAN-0196-01 {\sf Misgepop} and by
a grant from R\'egion Ile-de-France. The authors are grateful to the editorial team for its
encouraging comments.

\end{document}